\documentclass{article}
\usepackage{spconf,amsmath,graphicx,hyperref}
\usepackage{booktabs}

\usepackage[style=ieee, backend=bibtex, citestyle=numeric-comp, sorting=none, maxnames=6, minnames=1, giveninits=true, doi=false, isbn=false, url=false, eprint=false]{biblatex}

\defbibheading{bibliography}[\refname]{%
  \section*{\centering\MakeUppercase{#1}}%
}

\title{A SUPERB-Style Benchmark of Self-Supervised Speech Models for Audio Deepfake Detection}
\name{Hashim Ali, Nithin Sai Adupa, Surya Subramani, Hafiz Malik}
\address{Electrical and Computer Engineering\\
        University of Michigan\\
	4901 Evergreen Rd, Dearborn, MI 48128, USA}
\begin{document}
%
\maketitle
\begin{abstract}
Self-supervised learning (SSL) has transformed speech processing, with benchmarks such as SUPERB establishing fair comparisons across diverse downstream tasks. Despite it's security-critical importance, Audio deepfake detection has remained outside these efforts. In this work, we introduce Spoof-SUPERB, a benchmark for audio deepfake detection that systematically evaluates 20 SSL models spanning generative, discriminative, and spectrogram-based architectures. We evaluated these models on multiple in-domain and out-of-domain datasets. Our results reveal that large-scale discriminative models such as XLS-R, UniSpeech-SAT, and WavLM Large consistently outperform other models, benefiting from multilingual pretraining, speaker-aware objectives, and model scale. We further analyze the robustness of these models under acoustic degradations, showing that generative approaches degrade sharply, while discriminative models remain resilient. This benchmark establishes a reproducible baseline and provides practical insights into which SSL representations are most reliable for securing speech systems against audio deepfakes.
\end{abstract}
\begin{keywords}
speech self-supervised learning, audio deepfake detection, model generalization, benchmark
\end{keywords}
\section{Introduction}
\label{sec:intro}

The rapid advancement of text-to-speech (TTS) and voice conversion (VC) technologies has enabled the creation of highly realistic audio deepfakes. While these techniques offer opportunities for accessibility and content creation, their misuse threatens security, trust, and forensic reliability. Spoofing attacks have already been used to compromise automatic speaker verification (ASV) systems and spread misinformation. This makes robust audio deepfake detection an urgent research problem.

Self-supervised learning (SSL) has transformed the landscape of speech processing. Models such as wav2vec 2.0 \cite{baevski2020wav2vec}, HuBERT \cite{hsu2021hubert}, WavLM \cite{Chen2021WavLM}, and XLS-R \cite{babu2021xls} learn general-purpose representations from large unlabeled corpora and have achieved state-of-the-art performance on a wide range of tasks. To systematically evaluate SSL models, benchmarks such as SUPERB \cite{yang21c_interspeech}, SUPERB-SG \cite{tsai_superb-sg_2022}, MiniSUPERB \cite{wang_minisuperb_2023} and ML-SUPERB \cite{shi_ml-superb_2025} were introduced, which established standardized protocols for various downstream tasks. However, none of these benchmarks considers audio deepfake detection. A separate line of research has shown that SSL models are promising for antispoofing. Tak et al. \cite{tak2022automatic} demonstrated the effectiveness of wav2vec 2.0 for ASVspoof 2021 \cite{liu_asvspoof_2023}. Zhang et al. \cite{zhang_audio_2024} and Xiao and Das \cite{xiao_xlsr-mamba_2025} demonstrated state-of-the-art results on ASVspoof 21 DF and In-the-Wild datasets. While these works show a strong potential of SSL models for audio deepfake detection, existing studies remain fragmented across models, datasets, and protocols, making results difficult to compare.



In this work, we fill this gap by introducing the first SUPERB-style benchmark for audio deepfake detection. We evaluate a broad set of SSL models under a unified protocol: (i) frozen SSL front-ends with a weighted sum of all transformer layers, (ii) a simple fully connected classifier as the backend, and (iii) evaluation across multiple spoofing datasets, including ASVspoof 2019 (ASV19) \cite{wang2020asvspoof}, ASVspoof 2021 (ASV21) LA/DF \cite{liu_asvspoof_2023}, DeepfakeEval (DFEval) 2024 \cite{chandra2025deepfake}, In-the-Wild (ITW) \cite{muller_does_2022}, Famous Figures \cite{ali2025collecting}, and ASVSpoof Laundered Database (ASVSpoofLD) \cite{ali2024audio}. Our contributions are threefold:

\begin{enumerate}
    \item We present the first reproducible leaderboard of SSL models for audio deepfake detection.

    \item We analyze how architecture, pretraining data, and training objectives affect detection performance.

    \item We provide the first systematic comparison of SSL models under acoustically degraded conditions (e.g., codecs, reverberation, noise).
\end{enumerate}


\section{Related Work}
\label{sec:RW}

\subsection{SSL Benchmarks in Speech Processing}
\label{sub:superb}

The rise of self-supervised learning (SSL) has led to several large-scale benchmarks for general-purpose speech representations. The SUPERB framework \cite{yang21c_interspeech} standardized evaluation across tasks such as ASR, speaker recognition, and emotion recognition, and has since been extended: SUPERB-SG to semantic and generative tasks \cite{tsai_superb-sg_2022}, MiniSUPERB for lightweight evaluation \cite{wang_minisuperb_2023}, ML-SUPERB for multilingual modeling \cite{shi_ml-superb_2025}, and TS-SUPERB for target-speaker scenarios \cite{peng_ts-superb_2025}. These efforts established reproducible comparisons of SSL models, but none addressed audio deepfake detection.

\subsection{SSL for Audio Deepfake Detection}
\label{sub:ssl_antispoofing}
Parallel to benchmark development, a growing body of research has applied SSL models to audio deepfake detection. Tak et al. \cite{tak2022automatic} first showed that wav2vec 2.0 with fine-tuning and augmentation achieves strong results on ASVspoof 2021. The ASVspoof 5 Challenge \cite{wang24_asvspoof} then established SSL embeddings as the backbone of state-of-the-art systems: Rohdin et al. \cite{rohdin2024but} evaluated pooling strategies over SSL features, Chan et al. \cite{chan_enhancing_2024} fused WavLM-ResNet18-SA, and Chen et al. \cite{chen_ustc-kxdigit_2024} combined wav2vec2 embeddings with handcrafted features. The SAFE Challenge \cite{kirill2025safe} extended evaluation to multilingual and adversarial conditions, where the strongest systems again relied on SSL, such as mixture-of-experts frameworks \cite{negroni_leveraging_2025}, and WavLM+MAE-AST-Frame multilingual models \cite{ali2025multilingual}.

Beyond these challenges, recent research has developed specialized SSL frameworks. Zhang et al. introduced a Sensitive Layer Selection (SLS) classifier with XLS-R \cite{zhang_audio_2024}, achieving the first results below 2\% EER on the ASVspoof 2021 DF set and strong performance on In-the-Wild. Xiao and Das proposed XLSR-Mamba \cite{xiao_xlsr-mamba_2025}, combining wav2vec2 with a dual-column bidirectional state space model for efficient detection. Collectively, these works demonstrate that while SSL-based systems define the state of the art in audio deepfake detection, comparative insights remain fragmented due to differing datasets, backends, and evaluation protocols.



Despite progress, there is still no unified benchmark for audio deepfake detection analogous to SUPERB. Existing studies evaluate SSL models in isolation, making the results difficult to compare. Challenges like ASVspoof 5 and SAFE, and recent systems such as XLSR-SLS and XLSR-Mamba, all advanced performance but under inconsistent setups. What remains missing is a controlled analysis that reveals (i) which pretraining choices transfer best, (ii) how model design impacts detection, and (iii) which SSL representations resist acoustic degradations. To address this, we introduce a SUPERB-style benchmark for audio deepfake detection, providing reproducible evaluation of a broad family of SSL models under a unified protocol.


\section{Methodology: Spoof-SUPERB}
\label{sec:method}

In this section, we detail the \textbf{Spoof-SUPERB} framework, which adapts the SUPERB benchmarking methodology \cite{yang21c_interspeech,tsai_superb-sg_2022,wang_minisuperb_2023} to the task of audio deepfake detection. Spoof-SUPERB standardizes training on ASVspoof 2019 \cite{wang2020asvspoof} and evaluates SSL models across multiple spoofing datasets, enabling fair and reproducible comparison of representations.

\subsection{Self-Supervised Learning (SSL) Speech Models}
\label{sub:ssl_models}

\begin{table*}[t]
\centering
\footnotesize
\setlength{\tabcolsep}{3pt}
\renewcommand{\arraystretch}{0.95}
\caption{Details of investigated SSL representations. \textbf{Abbreviations:} F=Future, M=Masked, G=Generation, C=Contrastive, P=Token predic., VQ=Vector Quantization, LS=LibriSpeech, LL=LibriLight, GS=GigaSpeech, VP=VoxPopuli, AS=AudioSet, MLS=Multilingual LS, CV=CommonVoice, BBL=BABEL, G18=Googlei18n}
\label{tab:ssl_models_icassp}
\begin{tabular}{@{} l r r r r r r r @{}}
\toprule
\textbf{Model} & \textbf{Category} & \textbf{Network} & \textbf{\#Params} & \textbf{Stride} & \textbf{Input} & \textbf{Corpus} & \textbf{Pretraining} \\
\midrule
FBANK & Baseline & -- & -- & 10ms & waveform & -- & -- \\
\midrule
APC & Generative & 3-GRU & 4.11M & 10ms & FBANK & LS-360h & F-G \\
VQ-APC & Generative & 3-GRU & 4.63M & 10ms & FBANK & LS-360h & F-G + VQ \\
NPC & Generative & 4-Conv, 4-MaskedConv & 19.38M & 10ms & FBANK & LS-360h & M-G (+VQ) \\
Mockingjay & Generative & 12-Trans & $\sim$85M & 10ms & FBANK & LS-960h & time M-G \\
TERA & Generative & 3-Trans & $\sim$21M & 10ms & FBANK & LS-960h & time/freq M-G \\
DeCoAR 2.0 & Generative & 1D-Conv + 12-Trans & -- & 10ms & FBANK & LS-960h & time M-G (+VQ) \\
\midrule
wav2vec & Discriminative & 19-Conv & 32.54M & 10ms & waveform & LS-960h & F-C \\
wav2vec 2.0 Base & Discriminative & 7-Conv, 12-Trans & 95.04M & 20ms & waveform & LS-960h & M-C + VQ \\
wav2vec 2.0 Large & Discriminative & 7-Conv, 24-Trans & 317.38M & 20ms & waveform & LL-60kh & M-C + VQ \\
HuBERT Base & Discriminative & 7-Conv, 12-Trans & 94.68M & 20ms & waveform & LS-960h & M-P + VQ \\
HuBERT Large & Discriminative & 7-Conv, 24-Trans & 316.61M & 20ms & waveform & LL-60kh & M-P + VQ \\
MR-HuBERT & Discriminative & 7-Conv, 24-Trans & - & 20ms & waveform & LS-960h / LL-60kh & M-P + VQ (multi-res) \\
XLS-R & Discriminative & 7-Conv, 24-Trans & 300M & 20ms & waveform & VP+MLS+CV+BBL & M-C +VQ \\
UniSpeech-SAT & Discriminative & 7-Conv, 24-Trans & 316M & 20ms & waveform & LL-60kh+GS+VP & M-P + spk-adv. \\
Data2Vec & Discriminative & 7-Conv, 12-Trans & 95M & 20ms & waveform & LS-960h & EMA teacher regression \\
WAVLABLM & Discriminative & 7-Conv, 24-Trans & 316.61M & 20ms & waveform & CV+VP+MLS+G18 & M-P + denoise \\
WavLM Large & Discriminative & 7-Conv, 24-Trans & 317M & 20ms & waveform & LS-960h+GS+VP & M-P + denoise/mix \\
\midrule
SSAST & Hybrid & ViT (12-Trans) & 89M & -- & log-Mel & AS+LS-960h & patch M-G + C \\
MAE-AST-FRAME & Hybrid & ViT (12-Trans) & 89M & -- & log-Mel & AS+LS-960h & patch M-G + C \\
\bottomrule
\end{tabular}
\end{table*}


Self-supervised learning for speech has progressed rapidly, producing diverse model families with different pretraining objectives and scales. Following the SUPERB benchmark \cite{yang21c_interspeech}, we group SSL models into three broad categories based on their learning objectives: \textbf{generative}, \textbf{discriminative}, and \textbf{hybrid} models. This objective-driven taxonomy facilitates analysis of how different pretraining signals influence representation robustness in audio deepfake detection.

\textbf{Generative models:} These models learn representations by reconstructing portions of the input signal, such as predicting future frames or recovering masked segments. Early examples include APC (autoregressive predictive coding) \cite{chung2019unsupervised} and VQ-APC (APC with quantization) \cite{chung2020vqapc}. Mockingjay \cite{encoders2020mockingjay} and TERA \cite{liu2021tera} extend this by masking the spectrogram spans; DeCoAR 2.0 \cite{ling2020decoar} integrates bidirectional encoding with reconstruction; and NPC (non-autoregressive predictive coding) \cite{liu2020non} modifies masked convolutions. These models predate today’s dominant discriminative architectures.

\textbf{Discriminative models:} These models learn representations by distinguishing target samples (positive) from distractor samples (negative), typically through contrastive or predictive objectives. Wav2vec \cite{schneider2019wav2vec} and Modified CPC \cite{riviere2020unsupervised} introduced contrastive predictive coding. wav2vec 2.0 \cite{baevski2020wav2vec} improved this with masking and quantization, and large-scale successors such as HuBERT \cite{hsu2021hubert}, WavLM \cite{Chen2021WavLM}, UniSpeech-SAT \cite{chen2022unispeech}, Data2Vec \cite{baevski2022data2vec}, and XLS-R \cite{babu2021xls} expanded the paradigm with larger corpora, multilingual scaling, and training objectives such as teacher regression or adversarial speaker training.


\textbf{Hybrid models:} A parallel line of work combines masked reconstruction with discriminative learning objectives during pretraining. Models such as SSAST~\cite{gong2022ssast} and MAE-AST~\cite{baade2022mae} adopt patch-level masking and transformer-based architectures, while jointly leveraging reconstruction losses and discriminative objectives. The FBANK features are used as a non-SSL baseline for comparison.

\begin{table*}[t]
\centering
\caption{Equal Error Rate (EER, \%) of SSL models across evaluation datasets. For each dataset column, the \textbf{best-performing model} is shown in bold. In the \emph{Mean EER} column, the \textbf{top five models} are highlighted in bold.}
\label{tab:results_main}
\resizebox{\linewidth}{!}{
\begin{tabular}{lrrrrrrrrr}
\toprule
\textbf{Model} &
\textbf{ASV19 LA} &
\textbf{ASV21 LA} &
\textbf{ASV21 DF} &
\textbf{ASV5 Eval} &
\textbf{In-the-Wild} &
\textbf{DFEval 2024} &
\textbf{Famous Fig.} &
\textbf{ASVspoofLD} &
\textbf{Mean EER} \\
\midrule
FBANK & 42.828 & 43.155 & 44.789 & 49.838 & 48.393 & 47.113 & 48.427 & 47.672 & 46.527 \\
\midrule
APC & 10.075 & 16.335 & 22.276 & 33.311 & 36.889 & 42.662 & 58.402 & 34.345 & 31.787 \\
VQ-APC & 12.155 & 18.872 & 20.217 & 30.581 & 34.860 & 52.173 & 58.544 & 31.799 & 32.400 \\
NPC & 15.243 & 17.619 & 25.239 & 37.868 & 40.986 & 49.843 & 51.979 & 29.758 & 33.567 \\
Mockingjay & 15.430 & 19.798 & 25.312 & 40.217 & 35.848 & 49.800 & 40.975 & 56.033 & 35.427 \\
Mockingjay-960h & 13.801 & 25.525 & 22.584 & 37.866 & 52.387 & 52.130 & 49.953 & 59.283 & 39.191 \\
TERA & 9.112 & 26.572 & 17.254 & 35.656 & 39.894 & 54.251 & 49.282 & 57.565 & 36.198 \\
DeCoAR 2.0 & 7.628 & 12.352 & 18.990 & 29.571 & 35.029 & 49.800 & 54.452 & 22.126 & 28.743 \\
\midrule
wav2vec & 8.812 & 15.500 & 14.761 & 30.691 & 42.239 & 53.895 & 51.048 & 36.263 & 31.651 \\
wav2vec 2.0 Base & 4.661 & 11.452 & 10.046 & 18.698 & 40.945 & 56.981 & 51.921 & 32.891 & 28.449 \\
wav2vec 2.0 Large & 7.695 & 18.887 & 11.617 & 19.956 & 40.461 & 55.764 & 44.401 & 30.413 & 28.649 \\
HuBERT Base & 4.867 & 12.562 & 13.387 & 23.990 & 27.276 & 53.747 & 53.749 & 17.772 & 25.919 \\
HuBERT Large & 2.788 & 10.049 & 11.996 & 21.252 & 21.039 & 52.991 & 48.440 & 13.146 & \textbf{22.712} \\
MR-HuBERT & 2.478 & 9.074 & 11.635 & 23.056 & 23.799 & 49.696 & 52.720 & 11.645 & \textbf{23.006} \\
XLS-R & 1.985 & 14.096 & \textbf{4.314} & \textbf{14.394} & 20.073 & 45.392 & \textbf{29.598} & \textbf{9.420} & \textbf{17.409} \\
UniSpeech-SAT & \textbf{1.961} & \textbf{8.818} & 7.443 & 14.996 & \textbf{16.791} & 49.800 & 46.601 & 9.557 & \textbf{19.496} \\
Data2Vec & 7.695 & 11.877 & 16.511 & 26.773 & 29.249 & 50.808 & 53.092 & 16.418 & 26.678 \\
WAVLABLM & 3.631 & 15.380 & 9.847 & 21.115 & 23.402 & 52.530 & 52.660 & 15.500 & 24.258 \\
WavLM Large & 2.273 & 11.636 & 11.527 & 17.549 & 24.331 & 49.696 & 35.367 & 12.089 & \textbf{20.558} \\
\midrule
SSAST & 11.693 & 24.935 & 22.909 & 31.186 & 47.113 & \textbf{40.184} & 36.885 & 21.523 & 29.553 \\
MAE-AST-FRAME & 7.685 & 19.554 & 17.001 & 27.295 & 43.645 & 47.974 & 35.214 & 19.978 & 27.293 \\
\bottomrule
\end{tabular}}
\end{table*}

\subsection{Experimental Setup}
\label{sub:exp_setup}

Following SUPERB \cite{yang21c_interspeech}, we fix the parameters of the upstream SSL models for all downstream training. Frame-level hidden states from all layers are extracted for each utterance, and a trainable weighted-sum mechanism is used to aggregate them into contextualized representations. These representations are projected into a lower-dimensional space (256 units) and are pooled at the utterance level using mean pooling. The pooled vector is passed through a lightweight classifier consisting of linear layers with ReLU activation and dropout to produce binary spoof/bona-fide predictions. All experiments are conducted with a single run and a fixed random seed, ensuring comparability across SSL models.

All SSL models are trained on the ASV19 LA - train set \cite{wang2020asvspoof}, which has been widely adopted as a reference corpus in the antispoofing community. The objective of Spoof-SUPERB is not to optimize detection performance through training data selection, but to enable a fair and controlled comparison of SSL representations under a fixed downstream setup. Evaluation is conducted across multiple spoofing datasets, including ASV19 LA - eval \cite{wang2020asvspoof}, ASV21 LA/DF \cite{liu_asvspoof_2023}, ITW \cite{muller2022does}, DFEval 2024 \cite{chandra2025deepfake}, Famous Figures \cite{ali2025collecting}, and ASVSpoofLD \cite{ali2024audio}. This fixed training and cross-dataset evaluation protocol enables a controlled assessment of SSL generalization ability, with performance reported using equal error rate (EER). Although there is a potential overlap between the pretraining data of XLS-R and WavLM and ASVspoof 5, which is derived from MLS, these models demonstrate consistently strong performance across multiple evaluation datasets.

\section{Results and Analysis}
\label{sec:results}

\subsection{Overall Performance}
Table~\ref{tab:results_main} reports Equal Error Rates (EER) of all SSL models across evaluation datasets, with the FBANK baseline included for reference. SSL-based models consistently outperform FBANK, which yields a mean EER of 46.5\%. In contrast, several SSL systems reduce the mean EER below 25\%. Table \ref{tab:results_main} highlights the top 5 performing models in bold. These models are XLS-R (17.4\%), UniSpeech-SAT (19.5\%), WavLM Large (20.6\%), HuBERT Large (22.7\%), and MR-HuBERT (23.0\%). These results confirm the effectiveness of large-scale discriminative SSL models for audio deepfake detection. Earlier generative models such as APC (31.7\%), TERA (36.1\%), and Mockingjay (35.4\%) remain significantly behind discriminative approaches.

\subsection{Insights by Model Factors}
Among all SSL models, XLS-R and UniSpeech-SAT achieve the strongest overall performance (17.8\% and 19.5\% mean EER, respectively). We attribute the success of XLS-R to its large-scale multilingual pretraining on over 400k hours of speech across dozens of languages, which provides greater robustness under mismatched test conditions. UniSpeech-SAT uses a speaker-aware pre-training method that encourages the preservation of speaker identity, and proposes two approaches on top of HuBERT model. First, it proposes the utterance-wise contrastive learning, which  enhances single speaker information extraction to improve downstream tasks like speaker verification and speaker identification. Second, it proposes the utterance mixing augmentation, which benefits the multi-speaker tasks such as speaker diarization.

Top-performing models are consistently large-scale discriminative SSL models trained with masked prediction. XLS-R, UniSpeech-SAT, WavLM Large, MR-HuBERT, and HuBERT Large dominate the leaderboard, achieving mean EERs below 25\%. Model scale plays a central role, with large versions consistently outperforming their base counterparts. Multilingual pretraining (XLS-R, UniSpeech-SAT) provides further gains by exposing the models to broader phonetic and acoustic variability. Finally, advanced pretraining strategies, such as speaker-aware objectives in UniSpeech-SAT and masked region modeling in MR-HuBERT, strengthen representations for audio deepfake detection.

\subsection{Robustness under Acoustic Degradations}
To assess robustness, we evaluated all systems on a subset of the ASVspoofLD corpus \cite{ali2024audio}, which includes ASV19 LA audio files corrupted with babble noise (SNR = 10 dB) and reverberation (RT60 = 6s). The results in Table \ref{tab:results_main} (column ASVSpoofLD) reveal substantial variation in how SSL representations handle these degradations. The most resilient systems are XLS-R (9.4\%, +7.4\% from ASV19 LA), UniSpeech-SAT (9.5\%, +7.6\% from ASV19 LA), MR-HuBERT (11.6\%, +9.2\%), WavLM Large (12.1\%, +9.8\%), and HuBERT Large (13.1\%, +10.4\%). In contrast, generative systems such as TERA (57.6\%, +31\%) and Mockingjay-960h (59.3\%, +34\%) collapse under degradation. 

\begin{table}[t]
\centering
\vspace{-6pt}
\caption{Average EER (\%) across all codec conditions (ASV5 Eval). Representative models from each category}
\label{tab:codec_avg}
\begin{tabular}{lc}
\toprule
\textbf{Model} & \textbf{Avg. Codec EER} \\
\midrule
FBANK (Baseline)        & 49.8 \\
APC (Generative)        & 33.3 \\
XLS-R (Discriminative)  & 13.5 \\
UniSpeech-SAT (Discriminative) & 14.0 \\
WavLM Large (Discriminative)   & 18.1 \\
SSAST (Hybrid)           & 28.8 \\
\bottomrule
\end{tabular}
\end{table} 

We further analyze performance under Codec conditions (ASV5). Table \ref{tab:codec_avg} reports the average EER across all Codec conditions (C01--C11) for representative models. Discriminative SSL models (XLS-R, UniSpeech-SAT) again achieve the best robustness, with average EERs below 20\%.

\section{Conclusion} \label{conc}
In this work, we introduced Spoof-SUPERB, a SUPERB-style benchmark for audio deepfake detection. By evaluating 20 SSL models, we gained insight into which representations transfer most effectively to spoofing detection. Our results show that large-scale discriminative models such as XLS-R, UniSpeech-SAT, and WavLM Large consistently outperform generative approaches and remain more resilient under noise, reverberation, and codec conditions. Future work will extend this study by expanding the noise degradation analysis and examining SSL model performance across synthesis methods to better understand vulnerabilities to specific generation techniques, further advancing robust speech countermeasures against evolving deepfake threats.





\printbibliography

\end{document}